\documentclass[
    onecolumn,
    preprintnumbers,
    prd,
    showpacs
]{revtex4}
\usepackage{graphicx}%
\usepackage{amsmath}
\usepackage{subfigure,hyperref}
\usepackage[normalem,normalbf]{ulem}
\usepackage{amssymb}
\usepackage{bm}
\usepackage{enumerate}
\begin{document}
%-------------------------%-------------------------------------
\title{Density fluctuations in $\kappa$-deformed
     inflationary universe}
%------------------------------------------------------------
\author{Hyeong-Chan Kim}
\email{hckim@phya.yonsei.ac.kr}
%------------------------------------------------------------
\author{Jae Hyung Yee}%
\email{jhyee@phya.yonsei.ac.kr}
%-------------------------------------------------------------
\affiliation{Department of Physics, Yonsei University, Seoul
120-749, Korea.
}%
\author{Chaiho Rim}
\email{rim@chonbuk.ac.kr}
%-------------------------------------------------------------
\affiliation{Department of Physics, Chonbuk National University,
Chonju 561-756, Korea.
}%
%------------------------------------------------------------
\date{\today}%
%-------------------------------------------------------------------------
\bigskip
%-------------------------------------------------------------------------
\begin{abstract}
%-------------------------------------------------------------------------
\bigskip
We study the spectrum of metric fluctuation in $\kappa$-deformed
inflationary universe. We write the theory of scalar metric
fluctuations in the $\kappa-$deformed Robertson-Walker space, which
is represented as a non-local theory in the conventional
Robertson-Walker space. One important consequence of the deformation
is that the mode generation time is naturally determined by the
structure of the $\kappa-$deformation.

We expand the non-local action in $H^2/\kappa^2$, with $H$ being the
Hubble parameter and $\kappa$ the deformation parameter, and then
compute the power spectra of scalar metric fluctuations both for the
cases of exponential and power law inflations up to the first order
in $H^2/\kappa^2$. We show that the power spectra of the metric
fluctuation have non-trivial corrections on the time dependence and
on the momentum dependence compared to the commutative space
results. Especially for the power law inflation case, the power
spectrum for UV modes is weakly blue shifted early in the inflation
and its strength decreases in time. The power spectrum of far-IR
modes has cutoff proportional to $k^3$ which may explain the low CMB
quadrupole moment.
%-------------------------------------------------------------------------
\end{abstract}
%-------------------------------------------------------------------------
%-------------------------------------------------------------------------
%-------------------------------------------------------------------------
\maketitle
%-------------------------------------------------------------------------

%%%%%%%%%%%%%%%%%%%%%%%%%%%%%%%%%
\section{Introduction}
%%%%%%%%%%%%%%%%%%%%%%%%%%%%%%%%%%%
The history of the studies on the Cosmic Microwave Background (CMB)
anisotropies and on the cosmological fluctuations is closely linked
to that of the study of the standard cosmological
model~\cite{weinberg,kolb}. We now have high resolution maps of the
anisotropies in the temperature of the cosmic microwave
background~\cite{bennet}, and its accuracy of the resolution is
improving further. In relation to these observational data,
overviews of the theory of cosmological perturbation applied to
inflationary cosmology have been presented in
Refs.~\cite{branden,giovan}. The cosmological observations reveal
that the Universe has non-random fluctuations on all scales smaller
than the present Hubble radius. In the most currently studied models
of the very early universe it is assumed that the perturbations
originate from quantum vacuum fluctuations, which was first proposed
in a paper by Sakharov~\cite{sakharov}. With this, the inflationary
cosmology bears in it the `trans-Planckian problem": Since inflation
has to last for long enough time to solve several problems of
big-bang model and to provide a causal generation mechanism for CMB
fluctuations, the corresponding physical wavelength of these
fluctuations has to be smaller than the Planck length at the
beginning of the inflation~\cite{branden2}. Both of the theories of
gravity and of matter break down at the trans-Planckian scale. Many
methods have been proposed to cure the problem. The modification of
the dispersion relation, which was used to study the thermal
spectrum of black hole radiation~\cite{unruh}, was applied to
cosmology~\cite{martin}.  Modifications of the evolution of
cosmological fluctuations due to the string-motivated space-time
uncertainty relations, $\delta x_{\rm phys}\delta t \geq l_s^2$,
have been introduced by Brandenberger and Ho~\cite{ho}. It was shown
that the uncertainty relation plays a significant role in the
spectrum of the metric fluctuation~\cite{cai}. Greene et
al.~\cite{greene} proposed the initial states which give an
oscillatory contribution to the primordial power spectrum of
inflationary density perturbations. There have also been some
attempts to explain the low CMB quadrupole moment
contribution~\cite{piao} by using the pre-Big Bang scenario in
string theory~\cite{veneziano}.  The ambiguity of the action in the
presence of a minimal length cutoff in inflation by the boundary
terms are studied by Ashoorioon, Kempf, and Mann~\cite{asho}.
However, it is not easy to construct a consistent field theoretic
model which satisfies both the stringy space-time uncertainty
relation and the spatial homogeneity and isotropy of the
Robertson-Walker space. A direct non-commutative deformation of the
commutation relation,
\begin{eqnarray} \label{NCF}
[x^\mu, x^\nu]= i \theta^{\mu\nu},
\end{eqnarray}
introduces a preferred direction in space, which breaks the isotropy
of 3-dimensional space. Therefore, it would be interesting to
construct a space-time non-commutative theory which keeps the
spatial homogeneity and isotropy demanded by the Robertson-Walker
space-time.

Much attention has been given on the possibility of explaining the
observational data as a quantum gravitational effect. As a
theoretical framework to study these quantum gravity effects
phenomenologically ``Doubly Special Relativity" (DSR, also called
Deformed Special Relativity) was proposed by
Amelino-Camelia~\cite{giovanni}, where there exist two
relativistically invariant scales, the speed of light and the Planck
scale, and extensive studies have been
followed~\cite{girelli,okon,freidel,amelino,livine,amelino2}.
Recently, it was argued that the coordinate space of the DSR theory
defined in curved momentum space is described by the
$\kappa-$Minkowski space. Therefore, a good candidate to study the
quantum gravity effect to cosmology is to extend the
$\kappa-$Minkowski theory to $\kappa-$Robertson-Walker Space
($\kappa-$RWS) and to study the effect of the deformation in the
cosmological evolution. In $\kappa-$Minkowski space the space-time
coordinates are non-commuting generators of a quantized Minkowski
space-time. The $\kappa$-deformed Minkowski space-time introduces a
dimensionful quantum deformation parameter, $\kappa$, which can be
chosen to have the dimension of
mass~\cite{kosinski,glikman,daszkie}. A natural choice of this
deformation parameter is the Planck mass $\kappa=M_P$. It is
therefore important to construct consistent quantum field theoretic
framework in $\kappa-$Minkowski space, and explore the physical
effects~\cite{amelino} in cosmological evolution. In this respect,
Kowalski-Glikman~\cite{kowalski} have studied the effects on the
density fluctuations of the quantum $\kappa-$Poincar\'{e} algebra.

In Sec. II, we construct the theory of cosmological fluctuations in
$\kappa$-Deformed Inflationary Universe ($\kappa$-DIU). Starting
from the scalar-gravity theory in a flat Robertson-Walker space we
briefly summarize the linearized theory of scalar metric
fluctuations. After developing the $\kappa-$RWS, we write the theory
of scalar metric fluctuations in the deformed space. We show that
the scalar theory in $\kappa-$RWS space is described by a nonlocal
field theory in the conventional Robertson-Walker space. The
nonlocal action is series expanded in $H^2/\kappa^2$ in Sec. III,
where $H$ is the Hubble parameter, and is quantized. In Sec. IV and
V, we calculate the power spectra of scalar metric fluctuations for
the cases of exponential and power law inflations, respectively. We
show that the $\kappa-$deformation alters  both the time dependence
and the frequency dependence of the power spectrum nontrivially. In
Sec. V, we summarize our results.

%%%%%%%%%%%%%%%%%%%%%%%%%%%%%%%%%%%%%%%%%%%%%%%%%%%%%%%%%
\section{Density fluctuation in $\kappa$-deformed
    Robertson-Walker space-time}
%%%%%%%%%%%%%%%%%%%%%%%%%%%%%%%%%%%%%%%%%%%%%%%%%%%%%%%%%

If the Universe is quantum mechanically created with vacuum energy
dominance, it will inflate from the beginning. Even though there may
be other choices for the pre-inflationary universes, we assume that
the inflation starts from the beginning of the universe.

The calculations in this paper are carried out in 4-dimensional
spatially flat Robertson-Walker metric,
\begin{eqnarray} \label{metric}
ds^2=-d t^2+a^2(t)(d r^2+ r^2 d\Omega^2),
\end{eqnarray}
where $a(t)$ denotes the scale factor of expanding universe. For
later use, we introduce the conformal time $\eta$ defined by
\begin{eqnarray} \label{eta}
\eta=\int \frac{dt}{a(t)}.
\end{eqnarray}

The Einstein-Hilbert action for gravity coupled to scalar matter
field is
\begin{eqnarray} \label{S:sG}
S=\int d^4 x \sqrt{-g}\left[-\frac{R}{16\pi
G}+\frac{1}{2}\partial_\mu \varphi \partial^\mu
\varphi-V(\varphi)\right],
\end{eqnarray}
where $R$ is the Ricci curvature scalar. Starting from the
action~(\ref{S:sG}), it was shown that the scalar and tensor parts
of the linear metric fluctuation are described by the action (See
Refs.~\cite{ho,giovan}),
\begin{eqnarray} \label{S:s}
S=\frac{1}{2}\int_{\bf k} d\eta \left[(\partial_\eta{v_{\bf
    k}})^2+\left(\frac{\partial_\eta ^2 z}{z}-{\bf k}^2\right)
    v_{\bf k}^2 \right],
\end{eqnarray}
where $\displaystyle \int_{\bf k}\equiv \int \frac{d^3 {\bf
k}}{(2\pi)^3}$ and
\begin{eqnarray} \label{z}
z\equiv \frac{a(t)\dot \varphi_0}{H},
\end{eqnarray}
with $\varphi_0$ being the scalar zero mode and $H=\dot a/a$ the
Hubble parameter. It is noted that in cases of power law inflation
and of slow roll inflation, $H$ is proportional to $\dot \varphi_0$,
hence $z \propto a$. The action~(\ref{S:s}) can be cast into the
covariant form
\begin{eqnarray} \label{S:s2}
S&=&-\frac{1}{2}\int  d^4x \sqrt{-g}g^{\mu \nu}
    \partial_\mu {\cal R}\partial_\nu{\cal R},
\end{eqnarray}
where ${\cal R}$ is the gauge invariant metric fluctuation and plays
the role of a massless scalar field in the inflating universe. The
Fourier mode of ${\cal R}$, ${\cal R}_{\bf k}$, is related to the
field by $v_{\bf k}=z {\cal R}_{\bf k}$. Therefore, its power
spectrum is
\begin{eqnarray}\label{df:t}
P_{\bf k}=\frac{|{\bf k}|^3}{2\pi^2 z^2(t)} \langle 0|v_{\bf k}^2 |0
\rangle
 .
\end{eqnarray}

We now want to write the action~(\ref{S:s2}) in $\kappa-$RWS. For
this purpose, we develop scalar field theory in $\kappa-$RWS in the
following subsections. The $4$-dimensional field theory in
$\kappa-$Minkowski space has been constructed in
Ref.~\cite{amelino,girelli,kosinski} and references therein. We
briefly summarize the results obtained in these references in the
next subsection.

%%%%%%%%%%%%%%%%%%%%%%%%%%%%%%%%%%%%%%%%%%%%%%%%%%%%%%%%%%%
\subsection{ $\kappa$-deformed Minikowski space-time}
%%%%%%%%%%%%%%%%%%%%%%%%%%%%%%%%%%%%%%%%%%%%%%%%%%%%%%%%%%%%%
The $\kappa$-deformed Hopf algebra $H_x$ describing the
$\kappa-$Minkowski space is generated by the coordinates $\hat
x_\mu$ determined by the following relations:
\begin{eqnarray} \label{algebra}
~[\hat x_0, \hat x_i]&=& \frac{i}{\kappa} \hat x_i,~~ [\hat x_i,
\hat
   x_j]=0,~~\Delta(\hat x_\mu)=\hat x_\mu \otimes 1
    +1 \otimes \hat x_\mu .
\end{eqnarray}
The dual Hopf algebra $H_k$ of functions on $\kappa-$deformed
four-momenta is described by the Hopf subalgebra of the
$\kappa-$deformed Poincar\'{e} algebra as follows:
\begin{eqnarray} \label{algebra:p}
~[k_\mu, k_\nu]&=&0,~~\Delta(k_i)=k_i\otimes
e^{-k_0/\kappa}+1\otimes k_i,~~
    \Delta(k_0)=k_0\otimes 1+1\otimes k_0 .
\end{eqnarray}
The first Casimir operator of the algebra (\ref{algebra}) and
(\ref{algebra:p}) is
\begin{eqnarray} \label{Casimir}
M^2=\left(2 \kappa \sinh \frac{k_0}{2 \kappa }\right)^2- {\bf k}^2
e^{k_0/\kappa} .
\end{eqnarray}
It follows from this that for positive $\kappa$ the on-shell
three-momentum is bounded from above by
\begin{eqnarray} \label{klimit}
{\bf k}^2 \leq \kappa^2,
\end{eqnarray}
and the maximal value of momentum results in an infinite
energy~\cite{kow}.

By using $\kappa-$deformed Fourier transform, the fields on
$\kappa-$Minkowski space with non-commutative space-time coordinates
$\hat x=(\hat x_0,\hat x_i)$ is written as ($k\hat x\equiv k_i \hat
x_i-k_0 \hat x_0$):
\begin{eqnarray} \label{fourier}
\Phi(\hat x) = \int_k \tilde \Phi(k)
    :e^{i k \hat x}: ,
\end{eqnarray}
where $\displaystyle \int_k \equiv\int \frac{d^4k}{(2\pi)^4}$ and
$\tilde\Phi(k)$ is a classical function on commuting four-momentum
space $k=(k_0,k_i)$ and the normal ordering is defined by
\begin{eqnarray} \label{NOrdering}
:e^{i k \hat x}:\equiv e^{-i k_0 \hat x_0} e^{i \bf k \hat x}.
\end{eqnarray}
Multiplication of two normal ordered $\kappa-$deformed exponentials
follows from Eqs.~(\ref{algebra}) and (\ref{NOrdering}):
\begin{eqnarray} \label{normal ordering}
  :e^{i k \hat x}::e^{i q \hat x}:= :e^{i ({\bf k}
     e^{-q_0/\kappa}+ {\bf q}){\bf \hat x}-i (k_0+q_0) \hat x_0}: ,
\end{eqnarray}
which follows from the four momentum addition rule described by the
coproduct~(\ref{algebra:p}). From this we get the conjugate field,
\begin{eqnarray} \label{HC}
\Phi^\dagger(\hat x)&=& \int_k \tilde \Phi^\dagger (k) :e^{ik\hat
x}:,~~\tilde \Phi^\dagger({\bf k},k_0) = e^{3k_0/\kappa}
    \tilde \Phi^*(-e^{k_0/\kappa} {\bf k},-k_0 ).
\end{eqnarray}
The multiplication of fields can now be expressed as
\begin{eqnarray} \label{phi^2}
\int \Phi^2(\hat x) d^4x= \int_k \tilde\Phi(k)\tilde{\Phi}(-{\bf k}
e^{k_0/\kappa}, -k_0) .
\end{eqnarray}

The differential calculus and its covariance properties under the
action of $\kappa-$deformed Poincar\'{e} group have been constructed
in Ref.~\cite{kosinski,daszkie}. The left or right partial
derivatives $\hat\partial_A$ to define $\kappa$-deformed vector
field are given by
\begin{eqnarray} \label{derivative}
\hat \partial_A \Phi(\hat x) =
:\chi_A\left(\frac{1}{i}\partial_\mu\right) \Phi(\hat x) : ,
\end{eqnarray}
where  $\chi_A :e^{ik\hat x}:= :\chi_A(k_\mu) e^{ik\hat x}:$ and
\begin{eqnarray} \label{vector}
\chi_i(k_\mu)= e^{k_0/\kappa} k_i, ~~\chi_0=\kappa \sinh
\frac{k_0}{\kappa} + \frac{{\bf k}^2}{2\kappa} e^{k_0/\kappa}  .
\end{eqnarray}
The adjoint derivative $\hat
\partial_A^\dagger$ can be defined to satisfy
\begin{eqnarray} \label{cond:unamb}
\int d^4\hat x \Phi_1(\hat x) \hat
\partial_0 \Phi_2(\hat x)= \int d^4\hat x [\hat\partial_0^\dagger
\Phi_1(\hat x)]\Phi_2(\hat x),
\end{eqnarray}
which leads to
\begin{eqnarray} \label{dagger}
\hat \partial_\mu^\dagger \Phi(x) &=&
    \chi_\mu ^\dagger(\partial_\mu/i)\Phi(\hat x)
    ,
\end{eqnarray}
where
\begin{eqnarray} \label{..}
\chi_\mu^\dagger(k)=\chi_\mu(-e^{k_0/\kappa} {\bf k},-k_0).
\end{eqnarray}
Based on these, the $\lambda \phi^4$ field theory was constructed in
Ref.~\cite{kosinski}. In the next subsection we generalize the
formulation to the case of the Robertson-Walker space-time.

%%%%%%%%%%%%%%%%%%%%%%%%%%%%%%%%%%%%%%%%%%%%%%%%%%%%%%%%%%%%%%%%%
\subsection{The scalar field theory in $\kappa$-deformed Robertson
    Walker space-time}
%%%%%%%%%%%%%%%%%%%%%%%%%%%%%%%%%%%%%%%%%%%%%%%%%%%%%%%%%%%%%%%%%%%

To generalize the field theories in the $\kappa$-Minkowski space to
the curved space case, we must be careful in selecting the
coordinates which satisfy the commutation relation~(\ref{algebra}).
Since any non-decreasing reparametrization of $t$ is an equally good
time coordinate in commutative space-time, it is important to choose
the time coordinate for which the commutation relation,
\begin{eqnarray} \label{com:RW}
[\hat x_0, \hat x_i]=\frac{i}{\kappa} \hat x_i ,
\end{eqnarray}
is imposed. We note that a natural time coordinate consistent with
the commutation relation is the cosmological time $x_0=t$. This
choice ensures the same form of commutation relation satisfied by
the locally flat coordinates~$(\hat t, \hat X_i=a(\hat t) \hat
x_i)$: $[\hat t,\hat X_i]= a(\hat t) [\hat t, \hat x_i]= i \hat
X_i/\kappa$. This simplicity cannot be attained for other choices of
time coordinate. For example, consider a time coordinate $x_0$
defined by the $00$-part of the metric $g_{00}=-s^2(x_0)$. In the
locally flat coordinates, the commutator $[s(\hat x_0) \hat
x_0,a(\hat x_0) \hat x_i]=a(\hat x_0)s(\hat x_0)[\hat x_0, \hat
x_i]+a(\hat x_0)[s(\hat x_0), \hat x_i] \hat x_0$ is not simply
reduced to a well defined form of Eq.~(\ref{algebra}). In this sense
the natural choice for the time coordinate is the cosmological time
$t$ where $s(t)=1$. With the cosmological time $t$, the equations
(\ref{algebra})$\sim$(\ref{derivative}) can be used without
modification.

The generalization of $\kappa-$deformed vector fields in
$\kappa-$RWS can be written as Eq.~(\ref{derivative}) with the
operator $\chi_\mu$ defined by,
\begin{eqnarray} \label{vector}
\chi_i= e^{\partial_0/i\kappa} \frac{\partial_i}{i}, ~~\chi_0=\kappa
    \sinh \frac{\partial_0}{i\kappa} - \frac{1}{2\kappa}
   \partial_i e^{\partial_0/{2i\kappa}} g^{ij}
    \partial_je^{\partial_0/{2i\kappa}} ,
\end{eqnarray}
where the ordering of the time-dependent metric, $g^{ij}$, and
derivatives are determined by demanding the adjoint derivatives to
satisfy Eq.~(\ref{cond:unamb}), which gives
\begin{eqnarray} \label{chi:RW}
\chi_i^\dagger(k)&=&-k_i ,\\
\hat \partial_0^\dagger \Phi(\hat x)&=&-\kappa \sinh
\frac{\partial_0}{i\kappa} \Phi(\hat x)-\frac{1}{2\kappa}
e^{\partial_0/(2i\kappa)}\partial_i \left[ e^{\partial_0/(2i\kappa)}
\partial_j \Phi(\hat x) \right]g^{ij}(\hat t) .
\end{eqnarray}
In addition to Eq.~(\ref{cond:unamb}) we demand the condition
\begin{eqnarray} \label{consisit}
\hat
\partial_0^\dagger \Phi^\dagger(\hat x)=(\hat
\partial_0 \Phi(\hat x))^\dagger ,
\end{eqnarray}
to determine $\hat \partial_0^\dagger$.

Given the covariant derivatives and its adjoint derivatives, we can
write the action of a massless scalar field in $\kappa-$RWS as
\begin{eqnarray} \label{S}
S&=&-\frac{1}{2}\int  d^4x(\hat
\partial_\mu^\dagger \Phi^\dagger(\hat
    x)) \sqrt{-g} g^{\mu \nu}(\hat t)
    \hat \partial_\nu\Phi(\hat x)  \\
&=&\frac{1}{2}\int d^4x \left[(\hat\partial_0^\dagger \Phi^\dagger(
    \hat x)) a^3(\hat t)
    \hat\partial_0\Phi(\hat x)-(\hat\partial_i^\dagger \Phi^\dagger(
    \hat x)) a(\hat t)
    \hat\partial_i\Phi(\hat x)\right], \nonumber
\end{eqnarray}
where we choose the symmetric form in the action so that the metric
dependent factor is placed in the middle of the operator products.
It turns out that this choice gives the simplest form of the
interaction between different modes. In addition we do not consider
the change of measure~\cite{moller} due to the complication of the
non-commutative multiplication since what we are interested in in
this paper is to understand the main feature of the
$\kappa-$deformation on the metric fluctuation.

Using Eqs.~(\ref{fourier}), (\ref{normal ordering}), (\ref{HC}),
(\ref{vector}), and (\ref{chi:RW}) we obtain for the action
\begin{eqnarray} \label{S:nl}
S &=&\frac{\kappa^2}{4}\int_{\bf k}dt\left\{ g({\bf k},t)\tilde
\Phi_{-\bf k}(t-
    \frac{i}{\kappa})\tilde \Phi_{\bf k}(t+\frac{i}{\kappa})
       -a^3(t)\left[\rho_3(t)+g_3(t) \bar k^2(t)\right]
     \tilde \Phi_{-\bf k}(t)
        \tilde \Phi_{\bf k}(t)\right\} ,
\end{eqnarray}
where $\bar k$ denotes the relative ratio between the physical
momentum ($|{\bf k}|/a$) and the non-commutative scale
($\displaystyle \bar k(t)=\frac{|{\bf k}|}{a(t) \kappa}$) and the
coefficients $g({\bf k},t)$, $g_i(t)$ and $\rho_n(t)$ are given by
\begin{eqnarray} \label{rhos}
g({\bf k},t) &=& a^3(t)\left[1+(g_{1}(t)-2)\bar k^2
    +\frac{g_2(t)}{2}\bar k^4\right]
    , \\
g_1(t)&=& \frac{1}{2}\left[\frac{a^2(t)}{a^2(t+i/(2\kappa))}
    +\frac{a^2(t)}{a^2(t-i/(2\kappa))}\right]
        ,  \nonumber \\
g_2(t)&=& \frac{a^4(t)}{a^2(t+i/(2\kappa))a^2(t-i/(2\kappa))}
   ,   \nonumber \\
g_3(t) &=& \frac{1}{2}\left[
    \frac{a^3(t-i/\kappa)}{a(t)a^{2}(t-i/(2\kappa))}+
     \frac{a^3(t+i/\kappa)}{a(t)a^{2}(t+i/(2\kappa))}\right]
    , \nonumber \\
\rho_n(t)&=&
    \frac{1}{2}\left[\frac{a^n(t-i/\kappa)}{a^n(t)}
        +\frac{a^n(t+i/\kappa)}{a^n(t)}\right]
    .\nonumber
\end{eqnarray}
Note that the left hand sides of Eq.~(\ref{rhos}) are defined to
satisfy $\displaystyle \lim_{\kappa t\rightarrow \infty} g_i= 1=
\lim_{\kappa t\rightarrow \infty} \rho_n$. The action~(\ref{S:nl})
is highly non-local in that the fields are non-locally multiplied in
the action, in addition to the nonlocal coupling between the
background metric and the field modes. Each mode of the field,
$\tilde \Phi_{\pm \bf k}$, is diagonalized so that it is not coupled
to other modes of different $\bf k$.

%%%%%%%%%%%%%%%%%%%%%%%%%%%%%%%%%%%%%%%%%
\section{First order approximation and the Hamiltonian formulation}
%%%%%%%%%%%%%%%%%%%%%%%%%%%%%%%%%%%%%%%%%%%%%%%%%%

It is not possible to solve the nonlocal equation of motion derived
from the action~(\ref{S:nl}) exactly. The canonical formalism for
Lagrangians with non-locality of finite extent has been proposed by
Woodard~\cite{woodard}. However, we do not follow the formalism
since our purpose is to obtain information on how the
non-commutativity~(\ref{com:RW}) affects the evolution of the metric
fluctuation in inflationary Universe in a simple calculable form.
Instead, we use a perturbative expansion in the parameter $\bar
H^2\equiv H^2/\kappa^2$ and construct the Hamiltonian for the action
up to the first order in $\bar H^2$.

To have an approximation of the action~(\ref{S:nl}), we expand the
integrand in $\bar H^2$ as
\begin{eqnarray} \label{fPhi:red}
g(t)\Phi(t+\frac{i}{\kappa})\Phi(t-\frac{i}{\kappa})&=&g_S^{(0)}(t)
\Phi(t)^2+\frac{2 g_A^{(-1)}(t)}{\kappa^2} \dot
\Phi^2(t)+\left[6(g_S^{(-4)}-g^{(-4)}) \right. \\
&&\left.+\kappa^{-2}\left(8g_A^{(-3)}
    -4 g_S^{(-2)}-g^{(-2)}\right)-\frac{g}{4\kappa^4}
        \right] \ddot \Phi^2(t)+O(\bar H^6
    ) \nonumber,
\end{eqnarray}
where $g^{(-n)}(t)$ denotes the $n^{\rm th}$ indefinite integrals of
$g(t)$, and
\begin{eqnarray} \label{sa}
g^{(-n)}_S(t)\equiv
    \frac{1}{2}[g^{(-n)}(t+i/\kappa)+g^{(-n)}(t-i/\kappa)],
~~g^{(-n)}_A(t)\equiv\frac{\kappa}{2i}\left[g^{(-n)}(t+i/\kappa)
    -g^{(-n)}(t-i/\kappa)\right]
.
\end{eqnarray}

From Eqs~(\ref{S:nl}) and (\ref{fPhi:red}), we get the action up to
the order $O(\bar H^2)$,
\begin{eqnarray} \label{S:2}
S&=&\frac{1}{2}\int_{\bf k}dt~ a^3(t)\left\{ \mu({\bf k},t)
    \dot{\tilde \Phi}_{\bf k}(t) \dot{\tilde\Phi}_{-\bf k}(t)
        - \frac{{\bf k}^2}{a^2(t)}\left(g_4-
       \frac{g_5 \bar k^2}{4}
    \right) \tilde\Phi_{\bf k} \tilde\Phi_{-\bf k}
  + \frac{\gamma({\bf k},t)}{3\kappa^2}\ddot{\tilde\Phi}_{\bf k}
    \ddot{\tilde\Phi}_{-\bf k} + \cdots
    \right\} ,
\end{eqnarray}
where the coefficients are given by
\begin{eqnarray} \label{coef}
\mu({\bf k},t) &=&a^{-3}(t) g_A^{(-1)}({\bf k},t),\\
g_4(t) &=& \rho_1+\frac{1}{4}\left\{g_3- \frac{1}{2}
    \left[\frac{a^3(t+i/\kappa)}{a(t)a^2(t+3i/(2\kappa))}+
    \frac{a^3(t-i/\kappa)}{a(t)a^2(t-3i/(2\kappa))}
    \right]\right\},\nonumber \\
g_5(t)
&=&\frac{1}{2}\left[\frac{a(t)a^3(t+i/\kappa)}{a^2(t+i/2\kappa)
            a^2(t+3i/(2\kappa))}+
    \frac{a(t)a^3(t-i/\kappa)}{a^2(t-i/2\kappa)
            a^2(t-3i/(2\kappa))}
    \right], \nonumber \\
\gamma({\bf k},t)&=& \frac{3 }{2a^3(t)}
    \left[-6\kappa^4(g^{(-4)}-g_S^{(-4)})+\kappa^{2}\left(8g_A^{(-3)}
    -4 g_S^{(-2)}-g^{(-2)}\right)-\frac{g}{4}\right] .\nonumber
\end{eqnarray}
The asymptotic values of these coefficient functions are
\begin{eqnarray} \label{g,m:asym}
g_4(\infty)=g_5(\infty)=\mu(0,\infty)=\gamma(0,\infty)=1 ,
\end{eqnarray}
which make it easier to guess the asymptotic behaviors of the
coefficient functions for large $t$.

For notational simplicity, we use the change of variables
$\Phi_{{\bf k},+}=\frac{1}{2}(\tilde\Phi_{\bf k}+\tilde\Phi_{-\bf
k}),~~ \Phi_{{\bf k},-}=\frac{i}{2}(\tilde\Phi_{\bf
k}-\tilde\Phi_{-\bf k})$, to write the action in a diagonal form:
\begin{eqnarray} \label{S:+-}
S&=&\frac{1}{2}\int_{\alpha}dt ~a^3(t)\left\{\mu({\bf k},t)
    \dot\Phi^2_{\alpha}(t) - \frac{{\bf k}^2}{a^2}
        \left(g_4-\frac{g_5 \bar k^2}{4}
   \right) \Phi^2_{\alpha}
  +
  \frac{\gamma({\bf k},t)}{3\kappa^2}\ddot{\Phi}^2_{\alpha}+\cdots\right\}
  ,
\end{eqnarray}
where $\alpha = ({\bf k}, \pm)$. Note that the coefficients $\mu$,
$\gamma$, and $g_i$ are exactly calculable once $a(t)$ is given.

Introducing the conformal time $\eta$ and the rescaling of the
field, $\phi_{\alpha}=a(t(\eta))\Phi_\alpha$, we reduce the
action~(\ref{S:+-}) into the form, up to the order $\bar H^2$,
\begin{eqnarray} \label{S:+-2}
S&=&\frac{1}{2}\int_{\alpha}d\eta \left\{\bar \mu
    (\partial_\eta \phi_{\alpha})^2 - \omega({\bf k},t)\phi^2_{\alpha}
  + \frac{\gamma({\bf k},t)}{3\kappa^2a^2(t)}
    (\partial_\eta^2\phi_{\alpha})^2
  \right\},
\end{eqnarray}
where
\begin{eqnarray} \label{omega}
\bar \mu &=& \mu +\frac{4\gamma H^2}{3\kappa^2}\left(1+\frac{5\dot
    H}{4H^2}+\frac{3\dot\gamma}{4H\gamma}\right) ,\\
\omega({\bf k},t)&=&k^2\left(g_4-\frac{g_5\bar k^2}{
            4}
    \right)-\frac{\partial_\eta(A^2H)}{a}+\frac{\partial_\eta^2(
        \gamma \partial_\eta H)}{3 \kappa^2 a}, \\
A^2&=& a^2(\eta)\left[\mu({\bf k},t)+\frac{\partial_\eta(\gamma
 \partial_\eta a)}{3 \kappa^2 a^3}\right] . \nonumber
\end{eqnarray}
Note that the action~(\ref{S:+-2}) contains a higher derivative
term, which may lead to nonunitary evolution of the system. Since
this higher derivative term is a term of order $\bar H^2$ and our
purpose in this paper is to obtain the effect of the deformation on
the cosmological evolution up to the $1^{\rm st}$ order in $\bar
H^2$, we require that the higher order derivative term is written as
a function of the field, its first time derivative, and time:
\begin{eqnarray} \label{ass:H}
\partial_\eta^2\phi= \Psi(\phi, \partial_\eta \phi, \eta) .
\end{eqnarray}
Explicitly, we use the linearized approximation,
\begin{eqnarray} \label{ddphi:01}
\partial_\eta^2\phi= \frac{a(\eta)}{\gamma^{1/2}({\bf k},\eta)}
    [c(\eta) \phi +d(\eta) \partial_\eta\phi],
\end{eqnarray}
where the coefficients $c$ and $d$ are to be determined by
consistency.

This requirement is equivalent to the perturbative calculation up to
the 1$^{st}$ order in $\bar H^2$. This can be shown as follows: The
equation of motion for $\phi=\phi_0+ \frac{1}{\kappa^2}
\phi_1+\cdots $ can be written as
\begin{eqnarray} \label{eom0,1}
\partial_\eta (\bar \mu \partial_\eta \phi_0)+\omega \phi_0 &=&0, \\
\partial_\eta (\bar \mu \partial_\eta \phi)+\omega \phi&=&
    \frac{1}{3\kappa^2} \partial_\eta^2\left(\frac{\gamma}{a^2}
        \partial_\eta^2 \phi_0\right) , \nonumber
\end{eqnarray}
where the first equation is the 0$^{\rm th}$ order equation and the
second is the full equation written explicitly up to the 1$^{\rm
st}$ order. The first equation of Eq.~(\ref{eom0,1}) defines
$\partial_\eta^2 \phi_0$ as a linear function of $\partial_\eta
\phi_0$ and $\phi_0$. Then, the second line can be understood as a
defining equation of $\partial_\eta^2\phi$ as a linear function of
$\partial_\eta \phi$ and $\phi$ up to $O(\bar H^2)$, which is
Eq.~(\ref{ddphi:01}).

With this reasoning and Eq.~(\ref{ddphi:01}), the
action~(\ref{S:+-2}) is perturbatively equivalent up to $\bar H^2$
to the following unitary action
\begin{eqnarray} \label{S:fi}
S=\frac{1}{2} \int_{\alpha} d\eta \left[m (\partial_\eta
\phi_{\alpha})^{2}-
     f(k,\eta) \phi_{\alpha}^2\right],
\end{eqnarray}
where $k=|{\bf k}|$ and we have
\begin{eqnarray} \label{bmu0}
m&=& \bar \mu +\frac{d^2}{3\kappa^2},\\
f &=&\omega({\bf k},t)-\frac{c^2-\partial_\eta(c d)}{3\kappa^2}.
 \nonumber
\end{eqnarray}
Substituting Eq.~(\ref{ddphi:01}) into~(\ref{S:+-2})and requiring
the resultant action to be the same as the action (\ref{S:fi})
with~(\ref{ddphi:01}) as its equation of motion, we find
\begin{eqnarray} \label{cd}
c=-\frac{f\gamma^{1/2}}{m a},~~d=-\frac{\dot{m} \gamma^{1/2}}{m } .
\end{eqnarray}
Eqs.~(\ref{bmu0}) and (\ref{cd}) imply that $m$ and $f$ satisfy the
following differential equations:
\begin{eqnarray} \label{bmu}
m&=& \bar\mu +\frac{\gamma}{3\kappa^2} \frac{(\dot{ m})^2}{m^2}
    ,\\
\frac{f^2}{3\kappa^2 m^2 a^2}+ f &=&\omega({\bf k},t)
        +\frac{a(t)}{3\kappa^2}\frac{d}{dt}\frac{f\gamma
                \dot{m}}{a m^2} . \nonumber
\end{eqnarray}
Note that these conditions make the action to be a functional of
$\partial_\eta \phi$ and $\phi$.  The time evolution for the
theory~(\ref{S:fi}) is unitary and quantum mechanically well
defined.

We introduce mode dependent conformal time $\eta_k$ by
\begin{eqnarray} \label{etak}
d\eta_k= m^{-1}(k,\eta) d\eta=\frac{dt}{a(t)m(k,t)}.
\end{eqnarray}
Then the action~(\ref{S:fi}) can be written in a simplified form:
\begin{eqnarray} \label{S:final}
S=\frac{1}{2} \int_{\alpha} d\eta_k \left[\phi_{\alpha}'^{2}-
     \Omega^2_k(\eta_k) \phi_{\alpha}^2\right],
\end{eqnarray}
where $'$ denotes the derivative with respect to $\eta_k$ and
\begin{eqnarray} \label{Om:mf}
\Omega_k^2(\eta_k)\equiv m(k,\eta_k) f(k,\eta_k) .
\end{eqnarray}
The Hamiltonian for mode $\alpha$ is the same as that of the
time-dependent harmonic oscillator with frequency squared
$\Omega_k^2$:
\begin{eqnarray} \label{H}
H_\alpha= \frac{\hat\pi_\alpha^2}{2}+ \frac{1}{2} \Omega_k^2(\eta_k)
\hat \phi_{\alpha}^2 .
\end{eqnarray}
The time evolution of each mode can be described by introducing
invariant creation and annihilation operators~\cite{lvn},
\begin{eqnarray} \label{Adag}
\hat A_\alpha=-\frac{i}{\hbar^{1/2}} (\varphi_\alpha^* \hat
    \pi_\alpha-{\varphi_\alpha^*}' \hat \phi_\alpha ),~~
\hat A_\alpha^\dagger=\frac{i}{\hbar^{1/2}} (\varphi_\alpha \hat
\pi_\alpha-{\varphi_\alpha}' \hat \phi_\alpha ) ,
\end{eqnarray}
where $\varphi_\alpha$ is the mode solution of the differential
equation~(\ref{diff}) below and $\hat A_\alpha$ and $\hat
A_\alpha^\dagger$ satisfy the Liouville-von Neumann equation,
\begin{eqnarray} \label{Lvn}
i\hbar \partial_{\eta_k} \hat A_\alpha +[\hat A_\alpha, H_\alpha]=0
.
\end{eqnarray}
One may invert Eq.~(\ref{Adag}) to construct the field operator in
terms of the creation and annihilation operators as
\begin{eqnarray} \label{inv:fi}
\hat \phi_\alpha &=& \hbar^{1/2} \left[\varphi_\alpha(\eta_k) \hat
A_\alpha
    +\varphi_\alpha^*(\eta_k) \hat A_\alpha^\dagger \right],\\
\hat \pi_\alpha &=& \hbar^{1/2} \left[\varphi_\alpha'(\eta_k) \hat
A_\alpha
    +{\varphi_\alpha^*}'(\eta_k) \hat A_\alpha^\dagger \right].\nonumber
\end{eqnarray}
Note also that the Liouville-von Neumann equation is equivalent to
the following differential equation for the coefficients
$\varphi_\alpha$,
\begin{eqnarray} \label{diff}
\varphi_\alpha''(\eta_k) +\Omega^2_k(\eta_k)
\varphi_\alpha(\eta_k)=0 .
\end{eqnarray}
The commutation relation $[\hat A_\alpha, \hat A_\beta^\dagger
]=\delta_{\alpha\beta}$ restricts the mode solution $\varphi_\alpha$
to satisfy $\varphi_\alpha {\varphi_\alpha^*}'-\varphi_\alpha'
\varphi_\alpha^*=i$.

We present the first order approximation of $m$ and $f$ for later
use. To first order in $1/\kappa^2$, Eq.~(\ref{bmu}) gives
\begin{eqnarray} \label{dm}
m&\simeq & \bar\mu +\frac{\gamma}{3\kappa^2} \frac{(\dot{ \bar
\mu})^2}{\bar \mu^2}
    \simeq \bar \mu ,\\
f&=&2\omega({\bf k},t)\left[1+
    \left(1+\frac{4\omega}{3\kappa^2m^2a^2}\right)^{1/2}
    \right]^{-1}\simeq \omega
    \left(1-\frac{\omega}{3\kappa^2 m^2 a^2}\right). \nonumber
\end{eqnarray}
Using the explicit form for $\omega$ and $\mu$, we have
\begin{eqnarray} \label{f0}
m &\simeq& \bar \mu
 \simeq 1-\frac{\bar H^2}{6}(1-7\epsilon_1)
   -\bar k^2, \\
f&\simeq &k^2\left[1+\frac{\bar H^2}{3}(1-\epsilon_1)
    -\frac{7}{12}\bar k^2\right]-2 H^2 a^2\left[1
    +\frac{\epsilon_1}{2}-\frac{\bar H^2}{6}
    \left(1-16\epsilon_1-\epsilon_1^2+\frac{\epsilon_2}{2}
    +\epsilon_3\right )\right]
    ,
  \nonumber
\end{eqnarray}
where $\displaystyle \epsilon_n = \frac{H^{(n)}}{(H)^{n+1}}$, with
$\displaystyle H^{(n)}= \frac{d^n H}{dt^n}$, are constant numbers
for power law inflation and vanish for exponential inflation. From
these we have
\begin{eqnarray} \label{Om:gen}
m^{-1}(k,\eta) &\simeq & 1+
\alpha_n\frac{H^2}{\kappa^2}+\frac{k^2}{a^2
\kappa^2} +\cdots , \\
\Omega^2_k(\eta_k)& \simeq & k^2(1-w_1 \bar H^2-w_2 \bar k^2)-
    \left(2+\epsilon_1\right) H^2 a^2(\eta)(1-w_3\bar H^2)+\cdots .
    \nonumber
\end{eqnarray}
For the power law inflation, the values of $\alpha_n$ and $w_i$ are
given by
\begin{eqnarray} \label{ws}
\alpha_n=\frac{n+7}{6n},~~ w_1=\frac{1}{6}(13-11/n),~~w_2=19/12,~~
w_3=\frac{1}{3(1-1/(2n))}\left(1+\frac{45}{4n}-\frac{7}{4n^2}
-\frac{3}{n^3}\right) ,
\end{eqnarray}
and for the exponential inflation  their values are given by the
limits $n \rightarrow \infty$ of Eq.~(\ref{ws}).

%%%%%%%%%%%%%%%%%%%%%%%%%%%%%%%%%%%%%%%%%%%%%%%%%%%%%%%%%%%%%
\section{metric fluctuations in $\kappa-$DIU: The exponential
    inflation}
%%%%%%%%%%%%%%%%%%%%%%%%%%%%%%%%%%%%%%%%%%%%%%%%%%%%%%%%%%

The simplest inflationary model is the exponential inflation, in
which the scale factor $a(t)$ increases as,
\begin{eqnarray} \label{a:t}
a(t)= a_0 e^{H t} ,  ~~~-\infty < t <\infty .
\end{eqnarray}
Here $a_0$ is the scale factor at $t=0$ and $H$ is the Hubble
constant. Using the conformal time $\eta$, we get
\begin{eqnarray} \label{a}
Ht=-\ln (-a_0 H \eta),~~ a(\eta)=\frac{1}{-H\eta} ,
\end{eqnarray}
where the conformal time $\eta$  varies from $-\infty$ to $0$ as $t$
varies from $-\infty$ to $\infty$. From Eqs.~(\ref{coef}) and
(\ref{omega}), we have
\begin{eqnarray} \label{mu}
\bar \mu &=& \frac{\sin 3
    \bar H}{3 \bar H}+\frac{4}{3}\bar H^2
    \xi(3\bar H)
        + \left(-\frac{\sin \bar H}{\bar H}+\frac{2}{3}\bar H^2
            \xi(\bar H)
        \right)\left[2-\cos \bar H-\frac{\bar k^2}{2}\right]\bar
        k^2,\\
\gamma({\bf k},t)
    &=&\xi(3 \bar H)- \xi(\bar H)
    \left[2-\cos \bar H-\frac{\bar k^2}{2}\right]\bar
    k^2, \nonumber
\end{eqnarray}
where $\displaystyle \xi(x)=3\left[-3 \frac{1-\cos
x}{x^4}+4\frac{\sin x}{x^3}-\frac{4\cos
x+1}{2x^2}-\frac{1}{8}\right] \simeq 1-\frac{13}{80} x^2 +\cdots$.

Since the action~(\ref{S:final}) is obtained in the $\bar H^2$
expansion, the normalized Hubble constant, $\bar H=H/\kappa$ is
assumed to be smaller than one. Moreover, the Eq.~(\ref{k:max1})
below restricts $\bar k=k/(a(t)\kappa)$ to be smaller than one. Then
we get $m(k,\eta)$ from the differential equation~(\ref{bmu}) and
(\ref{mu}) by series expansion in $\bar H^2$ and $\bar k$,
\begin{eqnarray} \label{m:s}
m \simeq \bar \mu+\frac{\bar H^2}{3}\frac{\dot {\bar \mu}^2}{H^2\bar
\mu^2} \simeq \left(1-\frac{\bar H^2}{6}\right)(1-\bar H^2 k^2
\eta^2) +O(\bar H^4) .
\end{eqnarray}
By integrating $m^{-1}$ over $\eta$ using Eq.~(\ref{etak}), we get
the mode-dependent conformal time $\eta_k$,
\begin{eqnarray} \label{eta:eta}
\eta_k \simeq\frac{1}{2\bar H k(1- \bar H^2/6)}
    \ln \left(\frac{1+ \bar H k \eta}{1-
    \bar H k \eta}\right)
 = \frac{\eta}{1-\bar H^2/6}\left(1+\frac{\bar H^2}{3}
     k^2 \eta^2+\cdots \right),
\end{eqnarray}
which is normalized so that $\eta_k=0$ at $\eta=0$. A crucial point
is that there is a global rescaling of the conformal time due to the
non-commutative effect.

From Eq.~(\ref{Om:gen}), the effective frequency squared
$\Omega^2_k$ is written as
\begin{eqnarray} \label{Om:22}
\Omega^2_k(\eta_k)&= & k^2(1-w_1 \bar H^2-w_2 \bar k^2)-
    2H^2 a^2(\eta)(1-w_3 \bar H^2)+\cdots \\
&\simeq& \tilde k^2-\frac{\nu^2-1/4}{\eta_k^2}
        -w_2\bar H^2  k^4 \eta_k^2, \nonumber
\end{eqnarray}
where $w_1=13/6$, $w_2= 19/12$, $w_3=1/3$, and
\begin{eqnarray} \label{nu,k}
\tilde k^2 \equiv k^2\left[1-(w_1+2/3)\bar H^2\right],
~~\nu^2-1/4\equiv\frac{2(1-w_3 \bar H^2)}{(1-\bar H^2/6)^2}\simeq 2.
\end{eqnarray}
The explicit value of $\nu$ is $\nu=3/2$ to this order, which is the
same as the commutative space result. We note that both the
frequency and the mode solution have corrections from
non-commutativity for large $|\eta_k|$. The last term of the second
line in Eq.~(\ref{Om:22}) becomes negligible for $\eta_k\sim 0$.

%%%%%%%%%%%%%%%%%%%%%%%%%%%%%%%%%%%%%%%%%%%%%%%%%%%%%%%%%%%%
\subsection{Mode generation and the initial condition}
%%%%%%%%%%%%%%%%%%%%%%%%%%%%%%%%%%%%%%%%%%%%%%%%%%%%%%%%%%%%%%%

In the $\kappa-$RWS, spatial momentum is also restricted similarly
as in (\ref{klimit}). In terms of comoving momentum $k=|{\bf k}|$,
we have
\begin{eqnarray} \label{k:max0}
\frac{k}{a(t)} \leq \kappa .
\end{eqnarray}
This gives the upper bound of $k$
\begin{eqnarray} \label{k:max1}
k \leq k_{max}(t)\equiv \kappa a(t) .
\end{eqnarray}
The maximal value of the wave-number~(\ref{k:max1}) is very similar
to that used by Brandenberger and Ho~\cite{ho} except for the fact
that the maximal value used in Ref.~\cite{ho} is determined by an
effective scale factor modified by the Moyal star product in the
action.

Since $m$ is positive definite, the mode-dependent conformal time
$\eta_k$, Eq.~(\ref{etak}), is well defined and is an increasing
function of $t$. The relation~(\ref{k:max1}) implies that for a
given $k$, there exists a conformal time $\eta_k^0$, the time
saturating the relation~(\ref{k:max1}):
\begin{eqnarray} \label{eta:0}
a(t(\eta_k^0)) \equiv \frac{k}{\kappa}.
\end{eqnarray}
Then, the mode $\phi_k$ cannot exist before $\eta_k^0$. In other
words, $\eta_k^0$ is the generating time of the mode $\varphi_k$.
This provides a hint to one of the major issues in which state the
fluctuations are generated. To satisfy the continuity of the number
of quanta of the $k$ mode when the mode becomes physical at
$\eta_k^0$, it must be in the adiabatic vacuum state. This vacuum
state can be chosen to be the WKB mode solution,
\begin{eqnarray} \label{ini:ex}
\lim_{\eta_k \rightarrow \eta_k^0}\varphi_k(\eta_k)=\frac{1}{
    \sqrt{2 \Omega_k(\eta_k)}}\exp i\left(
     \int_{\eta_k^0}^{\eta_{k}}\Omega_k d\eta_k + \psi_k\right)
    ,
\end{eqnarray}
where the constant phase $\psi_k$ can be chosen conveniently.

%%%%%%%%%%%%%%%%%%%%%%%%%%%%%%%%%%%%%%%%%%
\subsection{inflationary evolution}
%%%%%%%%%%%%%%%%%%%%%%%%%%%%%%%%%%%%%%%%%%%%

The inflationary evolution of the mode solution is determined by
identifying $\Omega_k^2$ of Eq.~(\ref{Om:22}). The corresponding
commutative space values can be obtained by setting $\bar H=0$ in
Eq.~(\ref{Om:22}). With the $\Omega_k^2$ we have two different time
scales $\eta_k^c$ and $\eta_k^i$ defined by $\Omega'_k(\eta_k^c)=0$
and $\Omega^2_k(\eta_k^i)=0$, respectively. These time scales are
given by
\begin{eqnarray} \label{eta:bc}
\eta_k^c=-\frac{(\nu^2-1/4)^{1/4}}{k w_2^{1/4} \bar H^{1/2}} ,~~
\eta_k^i \simeq-\frac{(\nu^2-1/4)^{1/2}}{k}\left(1+
    \frac{(\nu^2-1/4)w_2\bar
    H^2}{2}+\cdots \right),
\end{eqnarray}
where $\Omega_k^2(\eta_k)$ increases while $\eta_k< \eta_k^c$ and
decreases later as shown in Fig. 1. It is positive definite when
$\eta_k< \eta_k^i$ and negative later. For $\eta_k> \eta_k^i$, the
modulus of $\phi_k$ increases in time. By Eq.~(\ref{eta:0}), the
mode $\phi_k$ is generated at $\eta_k^0\simeq -(k \bar H)^{-1}$.
Note that these three time scales satisfy
\begin{eqnarray} \label{3times}
\eta_k^0 \ll \eta_k^c\ll \eta_k^i ,
\end{eqnarray}
if $\bar H \ll 1$. Note also that during $\eta_k^0 < \eta_k
<\eta_k^c$, the condition for the WKB approximation,
\begin{eqnarray} \label{WKB:cond}
\frac{\partial_\eta\Omega_k^2}{\Omega_k^3}\sim 2 \omega_2 k
|\eta_k|\bar H^2 \ll 1,
\end{eqnarray}
is valid.
\begin{figure}[htbp]
\begin{center}
\includegraphics[width=.6\linewidth,origin=tl]{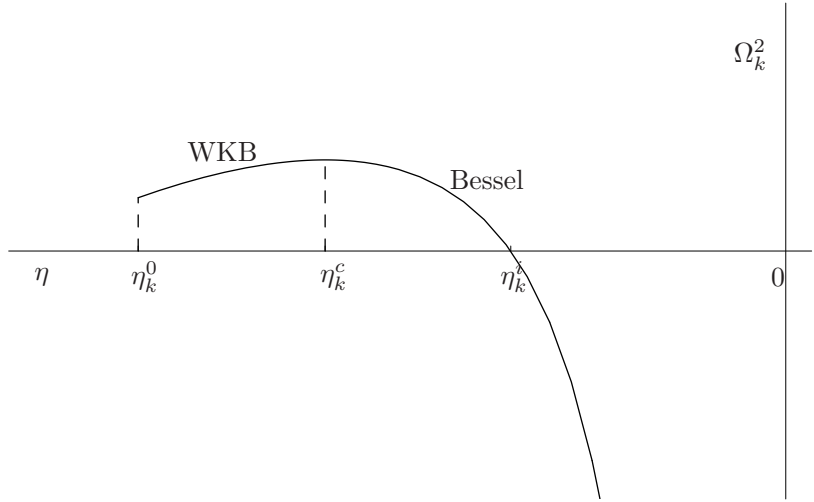}\hfill%
\end{center}
\caption{Schematic plot of $\Omega^2_k$ for the exponential
inflation and for the UV modes in the power law inflation. There is
no mode $\phi_k$ for $\eta_k<\eta_k^0$. The mode solution for
$\eta_k^0< \eta_k<\eta_k^c$ is given by the WKB solution and the
mode solution for $\eta_k> \eta_k^c$ is given by the Bessel
functions. The two solutions are matched at $\eta_k^c$.}
\label{o2:fig1}
\end{figure}
In Fig.~\ref{o2:fig1}, we present schematic plot of
$\Omega_k^2(\eta_k)$ for a given mode $\phi_k$. Therefore, during
this period, the WKB mode solution,
\begin{eqnarray} \label{sol:ini}
\varphi_k(\eta_k)=\frac{1}{
    \sqrt{2 \Omega(k,\eta_k)}}\exp i\left(
     \int_{\eta_k^0}^{\eta_{k}}\Omega_k d\eta_k +\psi_k\right),
\end{eqnarray}
can be used to describe the time evolution. We use this solution to
determine the matching condition at $\eta_k=\eta_k^c$,
\begin{eqnarray} \label{ini.}
\varphi_{k}(\eta_k^c)=\frac{1}{\sqrt{2\Omega_c}},~~~
    \varphi_k'(\eta_k^c)=i\sqrt{\frac{\Omega_c}{2}} ,
\end{eqnarray}
where $\Omega_c^2 \equiv \Omega^2_k(\eta_k^c)\simeq \tilde
k^2[1-2(\nu^2-1/4)^{1/2}\bar H ]$, $\tilde k$ given in (68), and
$\psi_k$ is chosen to give this matching condition~(\ref{ini.}).

For $\eta_k>\eta_k^c$, the last term in Eq.~(\ref{Om:22}) becomes
much smaller than other terms. Thus we ignore this term and use the
Bessel function as the solution for $\eta_k>\eta_k^c$,
\begin{eqnarray} \label{v:0}
\varphi_{k}(\eta_k)=A_k \sqrt{-\eta_k} J_\nu(-\tilde k \eta_k)+ B_k
\sqrt{-\eta_k} Y_\nu(-\tilde k\eta_k) .
\end{eqnarray}
Matching the two solutions at $\eta_k=\eta_k^c$, we get
\begin{eqnarray} \label{AB:L}
A_k&=&\frac{\pi}{2}
    \left[i \sqrt{-\eta_k^c}Y_\nu(-\tilde k\eta_k^c)
    \varphi_k'(\eta_k^c)
    +\frac{Y_\nu(-\tilde k\eta_k^c)
    -2\tilde k\eta_k^c Y_\nu'(-\tilde k\eta_k^c)}{
        2\sqrt{-\eta_k^c}} \varphi_k(\eta_k^c)\right], \\
B_k&=&-\frac{\pi}{2}
    \left[i\sqrt{-\eta_k^c}J_\nu(-\tilde k\eta_k^c)\varphi_k'(\eta_k^c)
    +\frac{J_\nu(-\tilde k\eta_k^c)
    -2\tilde k\eta_k^cJ_\nu'(-\tilde k\eta_k^c)}{
    2\sqrt{-\eta_k^c}}  \varphi_k(\eta_k^c)\right] .\nonumber
\end{eqnarray}
Note that $A_k$ and $B_k$ are independent of $k$ since $\tilde k
\eta_k^c$ is independent of $k$ due to Eq.~(\ref{eta:bc}) and
$\Omega_c \propto k$.

As $\-k\eta_k\rightarrow 0$, the Bessel functions become
\begin{eqnarray} \label{sol:0}
J_\nu \rightarrow \frac{(-\tilde k \eta_k)^{\nu}}{2^{\nu}
\nu!},~~Y_\nu \rightarrow \frac{2^{\nu}(\nu-1)!}{\pi(-\tilde k
\eta_k)^{\nu}} ,
\end{eqnarray}
and the second term in Eq.~(\ref{v:0}) dominates in the later time
($-k\eta_k \sim 0$). Therefore, the power spectrum of the scalar
metric perturbation has the form
\begin{eqnarray} \label{pS:EX}
P_{k}(t)= \frac{k^3}{2\pi^2}\frac{|\varphi_k(t)|^2}{z^2(t)}
    \simeq \frac{
    [2^\nu(\nu-1)!]^2|B|^2H^2}{2\pi^4(z/a)^2}
        \frac{(1+ \bar H^2/6)^2}{(1-17\bar H^2/6)^{3/2}}
        \left(\tilde k \eta_k\right)^{3-2\nu}
    ,
\end{eqnarray}
where $z/a$, $\bar H$, $\nu$, and $B$ are constant numbers. Note
that $3-2\nu \simeq \frac{2}{3}(w_3-2 \alpha) \bar H^2=0$ in the
present case since $w_3=1/3$. Therefore, the spectrum of the metric
fluctuation for exponential inflation in $\kappa-$RWS is time
independent and is scale invariant up to the first order in $\bar
H^2$. The only effect of the non-commutativity to the power spectrum
is the global rescaling of the power spectrum, which is of the order
$\bar H^2$. It is an interesting fact that $w_3$ and $2\alpha$ are
the same. Note that $\alpha$ originates from the scale factor of the
mode-dependent conformal time $\eta_k$ with respect to the conformal
time $\eta$, and $w_3$ comes from the $\bar H^2$ order correction
term of the frequency squared. Since there is no physical reason for
the coincidence, it is possible that the next order correction may
give a result of $w_3>1/3$. This is an interesting possibility since
this positive power of $k \eta$ makes the power spectrum decrease in
time. If the present analysis is applied to the tensor mode
fluctuation, the time dependence can be used to solve the
gravitational hierarchy problem ($H/M_P \sim
10^{-5}$)~\cite{rubakov}. This is what happens in the power law
inflation considered in the next section. The spectrum~(\ref{pS:EX})
is scale invariant in contrast to that of Ref.~\cite{kowalski} with
the same initial vacuum state. The difference may be attributed to
the different choice of the initial conditions. At the present case,
the initial time is dependent on the mode through Eq.~(\ref{eta:0}),
which uniquely fixes the initial state~(\ref{ini:ex}) for the mode
solutions.

%%%%%%%%%%%%%%%%%%%%%%%%%%%%%%%%%%%%%%%%%%%%%%%%%%%%%%%%%%%%%
\section{metric fluctuations in $\kappa-$DIU: Power law inflation}
%%%%%%%%%%%%%%%%%%%%%%%%%%%%%%%%%%%%%%%%%%%%%%%%%%%%%%%%%%

In this section we calculate the metric fluctuation in the power law
inflationary model, in which the scale factor increases as,
\begin{eqnarray} \label{a:t}
a(t)= a_0 (\kappa t)^n, ~~ 0<t<t_f,
\end{eqnarray}
where $n \neq 1$, $a_0$ is the scale factor at the Planck time
$t=1/\kappa$, and $t_f$ is the instance when the inflation ends. In
this model, the variable $z(t)$ in Eq.~(\ref{z}) is given by
$\displaystyle z(t)= \sqrt{\frac{2}{n}} M_P a(t)$. For $n \neq 1$,
we have
\begin{eqnarray} \label{a}
\kappa t=\left(\frac{\eta}{\eta_0}\right)^{\frac{3}{2}-\mu},~~
a(\eta)
   =a_0\left(\frac{\eta}{\eta_0}\right)^{\frac{1}{2}-\mu} ,
\end{eqnarray}
where $\displaystyle \mu=\frac{3n-1}{2(n-1)}$, and $\eta_0$ is the
conformal time corresponding to $\kappa t=1$, given by
\begin{eqnarray} \label{eta:I}
\eta_0=-\frac{\mu-3/2}{a_0\kappa} .
\end{eqnarray}
$\eta_f= \eta_0 (\kappa t_f)^{-1/(2\mu-3)}$ is the time when the
inflation ends.

When $\kappa t\gg 1$ is large and $k^2/(a^2\kappa^2) \ll 1$ is
small, we have
\begin{eqnarray} \label{m:s}
m^{-1}(k,\eta) \simeq 1+
\alpha_n\frac{H^2(\eta)}{\kappa^2}+\frac{k^2}{a^2(\eta) \kappa^2}
+\cdots.
\end{eqnarray}
The mode-dependent conformal time $\eta_k$ can be approximated as
\begin{eqnarray} \label{eta:eta}
\eta_k \simeq \eta\left[1
    +\frac{\alpha_n n^2}{2(\mu-1)} \bar \eta^{2\mu-3}
    +\frac{k^2}{2\mu\kappa^2 a_0^2}
    \bar \eta^{2\mu-1} +\cdots\right] ,
\end{eqnarray}
where we use the notation
\begin{eqnarray} \label{bar eta}
 \bar \eta \equiv \frac{\eta}{\eta_0}=-\frac{a_0 \kappa \eta}{\mu-3/2},
\end{eqnarray}
and we normalize the time so that $\eta_k=0$ when $\eta=0$.  The
function $\Omega_k^2$ in Eq.~(\ref{Om:gen}) for $\bar H^2 \ll 1$ is
approximated to be
\begin{eqnarray} \label{Om:pw1}
\Omega^2_k(\eta_k)&\simeq & k^2(1-w_1 H_k^2-w_2 \bar k^2)-
    \frac{2n-1}{n}H^2 a^2(\eta)(1-w_3h^2) +\delta ,
\end{eqnarray}
where $\delta$ represents the smaller terms proportional to the
differences of the Hubble parameter from its time-averaged values,
\begin{eqnarray} \label{delta}
\delta&=&-w_1k^2(\bar
    H^2-H_k^2)+2w_3H^2 a^2(\bar H^2-h^2).
\end{eqnarray}
$w_i$'s are given in Eq.~(57) and the time-averaged values of the
Hubble parameters, $H_k$ and $h$ are defined by
\begin{eqnarray} \label{Hk}
H_k^2 \equiv \frac{\int d\eta_k H^2(\eta)}{\kappa^2\int
d\eta_k},~~~~~
    h^2 \equiv  \frac{ \int d\eta_k a^2(\eta)
    H^4(\eta) }{\kappa^2\int d\eta_k  a^2(\eta) H^2(\eta) },
\end{eqnarray}
with the $\eta_k$ integrations performed over the validity range of
the differential equation~(\ref{diff}) for a given mode solution,
which will be clarified in the next subsections. We do not put the
index $k$ to $h$ since $h$ depends on $k$ very weakly.

%%%%%%%%%%%%%%%%%%%%%%%%%%%%%%%%%%%%%%%%%%%%%%%%%%%%%%%%%%%%
\subsection{Mode generation and the initial condition}
%%%%%%%%%%%%%%%%%%%%%%%%%%%%%%%%%%%%%%%%%%%%%%%%%%%%%%%%%%%%%%%

As in the case of the $\kappa-$deformed exponentially inflating
universe, due to the condition~(\ref{k:max1}), the mode $\phi_k$ is
generated at the conformal time $\eta(k)$,
\begin{eqnarray} \label{eta0:pw}
\eta(k)=\eta_0\left(\frac{k}{a_0 \kappa}\right)^{-\frac{1}{\mu-1/2}}
,
\end{eqnarray}
where we assume the Universe is inflating with $n \gg 1$.

A serious obstacle in finding physics of low comoving momentum modes
is that the action~(\ref{S:final}) is not well defined for large
$\bar H^2$ since the action is approximated by expansion in $\bar
H^2$. The condition $\bar H(\eta_m) \sim 1$ is attained at
\begin{eqnarray} \label{eta:1}
\eta_m \sim \eta_0 ~n^{-\frac{1}{2\mu-3}} .
\end{eqnarray}
Before $\eta_m$, our approximation for the action~(\ref{S:final}) is
not valid. This condition restricts the validity range of the
present approximation to the modes $\phi_k$ with comoving momentum
\begin{eqnarray} \label{valid}
k > k_m \equiv n^{\frac{\mu-1/2}{\mu-3/2}} a_0 \kappa .
\end{eqnarray}
In this subsection, we restrict ourselves to the ultra-violet modes
satisfying the condition~(\ref{valid}). To know the behavior of
smaller frequency modes, one should use better approximation of the
action~(\ref{S:nl}) instead of (\ref{S:final}). We present some
reasonable arguments for the evolution of those low frequency modes
in Sec. V.C.

For modes $k > k_m$, all the arguments for the initial state in Sec.
IV.A hold true. Therefore, the initial state is given by the WKB
ground state,
\begin{eqnarray} \label{ini:cond}
\varphi_k(\eta_k)=\frac{1}{
    \sqrt{2 \Omega_k(\eta_k)}}\exp i\left(
     \int_{\eta_k^0}^{\eta_{k}}\Omega_k(\eta_k) d\eta_k + \psi_k\right)
    , ~~ k\geq k_m,
\end{eqnarray}
where $\eta_k$ is close to the mode generation time $\eta_k^0$ given
by Eq.~(\ref{eta:0}).
\begin{figure}[htbp]
\begin{center}
\includegraphics[width=.6\linewidth,origin=tl]{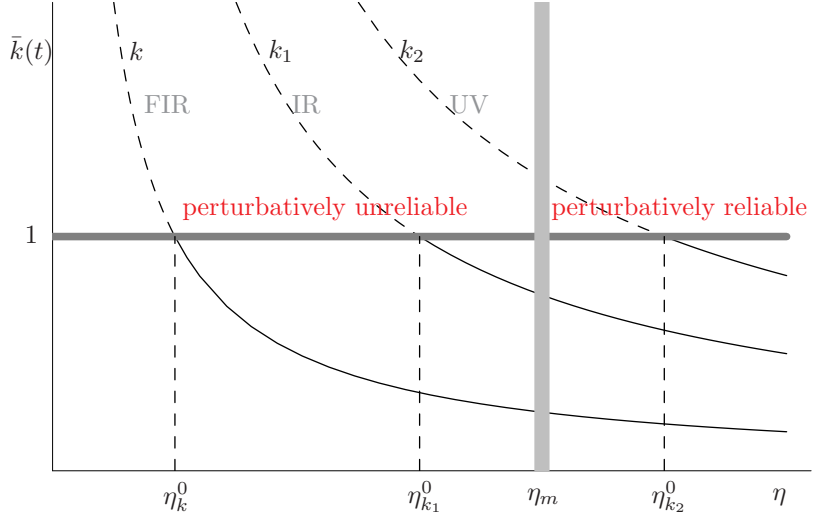}\hfill%
\end{center}
\caption{Schematic plot for the mode generation time in power law
inflation. Each curve describes the value of the normalized physical
wave-number $\bar k(t)=\frac{k}{a \kappa}$ for a given conformal
time $\eta$. Since $a(t(\eta))$ increases for expanding universe,
the value of $\bar k$ always decreases. Due to Eq.~(\ref{eta0:pw})
each mode becomes physical when it crosses the horizontal line $\bar
k=1$. We denote the unphysical part of each modes using the dashed
curve. Since $\bar H=1$ at $\eta=\eta_m$, the region $\eta> \eta_m$
is perturbatively reliable and the region $\eta< \eta_m$ is
perturbatively unreliable, which are divided by a shaded vertical
line. The UV modes resides in the perturbatively reliable region.
The IR and the Far Infra-Red ( $ \frac{k}{a_0 \kappa}<1$) modes pass
through the perturbatively unreliable region.
 } \label{o2:fig2}
\end{figure}

%%%%%%%%%%%%%%%%%%%%%%%%%%%%%%%%%%%%%%%%%%
\subsection{inflationary evolution}
%%%%%%%%%%%%%%%%%%%%%%%%%%%%%%%%%%%%%%%%%%%%
The inflationary evolution of the mode solution is governed by
$\Omega_k^2$. When $k>k_m$, we approximate $\Omega_k^2$ in
Eq.~(\ref{Om:pw1}) using Eqs.~(\ref{eta:eta}) and (\ref{delta}), and
dropping the term $\delta$ in Eq.~(\ref{Om:pw1}), as
\begin{eqnarray} \label{Om:pw}
\Omega^2_k(\eta_k)&\simeq & \tilde k^2-\frac{\nu^2-1/4}{\eta_k^2}
  - \frac{w_2 k^4}{\kappa^2 a_0^2}\bar \eta_k^{2\mu -1}+ \cdots.
\end{eqnarray}
where
\begin{eqnarray} \label{k nu:22}
\tilde k^2\equiv k^2\left[1-(w_1+\frac{2n-1}{\mu n})H_k^2
  \right],~~~
  \nu^2-\frac{1}{4}\equiv\left(\mu^2-\frac{1}{4}\right)
  \left[1+\left(\frac{\alpha_n}{\mu-1}-w_3\right)h^2 \right] .
\end{eqnarray}
Eq.~(\ref{Om:pw}) looks similar to Eq.~(\ref{Om:22}) except for the
final term. With the $\Omega^2_k(\eta_k)$ in Eq.~(\ref{Om:pw}), we
have two different time scales defined by $\Omega_k'(\eta_k^c)=0$
and $\Omega_k^2(\eta_k^i)=0$:
\begin{eqnarray} \label{etas:pw}
\eta_k^c
    &=&\eta_0 \left[\frac{\nu^2-1/4}{w_2(\mu-1/2)(\mu-3/2)^2\bar
k_0^4}\right]^{1/(2\mu+1)}
 ,\\
{\eta_k^i}^2&=&\frac{(\nu^2-1/4)}{\tilde k^2}
    \left[1+w_2(\mu-3/2)\left(\frac{\nu^2-1/4}{
    \mu-3/2}\right)^{2\mu}\bar k_0^{3-2\mu}\right]+\cdots \nonumber
\end{eqnarray}
where $\bar k_0= k/(\kappa a_0)$. Since $\bar k_0 \gg 1$ for modes
under consideration, we have
\begin{eqnarray} \label{etas}
\eta_k^0 \ll \eta_k^c \ll \eta_k^i .
\end{eqnarray}
Since the condition for the WKB approximation,
${\Omega_k^2}'/\Omega_k^3 \ll 1$, holds during $\eta_k <\eta_k^c$,
we may use the WKB solution~(\ref{ini:cond}). Therefore, we get the
mode solution at $\eta_k^c$ given by~Eq.(\ref{ini.}) with
\begin{eqnarray} \label{Omc}
\Omega_c^2\equiv \Omega_k^2(\eta_k^c)\simeq
    \tilde k^2\left[1-\frac{2\mu+1}{2\mu-1}
        \frac{\nu^2-1/4}{(\tilde k \eta_k^c)^2} \right] ,
\end{eqnarray}
where the second term in the parenthesis is smaller than 1 for $\bar
k_0 \gg 1$. For $\eta_k
> \eta_k^c$, we ignore the last term in Eq.~(\ref{Om:pw}) since
$\bar \eta_k$ is very small there. The solution for the differential
equation~(\ref{diff}) is given by the Bessel function in
Eq.~(\ref{v:0}) with parameters given in Eq.~(\ref{k nu:22}). With
the initial condition~(\ref{ini.}) and $\Omega_c$ in
Eq.~(\ref{Omc}), $A_k$ and $B_k$ are given by Eq.~(\ref{AB:L}).

Since we are considering modes with $\bar k \gg 1$, we always have
$|\tilde k \eta_k^c| \gg 1$. Using the asymptotic expansion of the
Bessel functions,
\begin{eqnarray} \label{vp:H}
J_{\nu}(x) \simeq \sqrt{\frac{2}{\pi x}} \cos [x
    -(\nu+\frac{1}{2})\frac{\pi}{2}] , ~~
Y_{\nu}(x) \simeq \sqrt{\frac{2}{\pi x}} \sin [x
    -(\nu+\frac{1}{2})\frac{\pi}{2}] ,
\end{eqnarray}
we get
\begin{eqnarray} \label{AB:UV}
A_k=\frac{\sqrt{\pi}}{2}\sqrt{\frac{\Omega_c}{\tilde k}}e^{-i(\tilde
k \eta_k^c+(\nu+1/2)\pi/2)}=i B_k .
\end{eqnarray}
Note that the absolute values of $A_k$ and $B_k$ are very weakly
dependent on $k$ since $\Omega_c\sim \tilde k$.

From this, we have the power spectrum,
\begin{eqnarray} \label{pS:UV}
P_{UV,k}(t)= \frac{k^3}{2\pi^2}\frac{|\varphi_k(t)|^2}{z^2(t)}
    \simeq \frac{\kappa^2}{M_P^2}
        \frac{n[2^{\nu}(\nu-1)!]^2(\mu-3/2)^{1-2\nu}}{
        16\pi^3 }
    \left(\frac{k}{a_0\kappa}\right)^{3-2\nu}\left(
    \frac{\eta_k}{\eta_0}\right)^{2(\mu-\nu)} +\cdots .
\end{eqnarray}
Since we are interested in the time evolution for $\eta_k^c <\eta_k
< \eta_f$, we have the time-averaged Hubble parameters,
\begin{eqnarray} \label{Hk:v}
H_k^2 &\simeq& \frac{n^2}{2(2-\mu)}\frac{(\bar
    \eta_f)^{4-2\mu}
        -(\bar \eta_k^c)^{4-2\mu}}{\bar \eta_f
        -\bar \eta_k^c}
    \simeq\frac{n^2}{2(2-\mu)}\frac{1}{(\bar \eta_k^c)^{2\mu-3}} ,\\
h^2 &\simeq & \frac{n^2}{2(2-\mu)} \frac{\bar
     \eta_f^{2\mu-4} - (\bar \eta_k^i)^{2\mu-4} }{\bar \eta_f^{-1}
     -(\bar \eta_k^i)^{-1} }
    \simeq \frac{n^2}{2(2-\mu)} (\bar
     \eta_f)^{2\mu -3} .
\end{eqnarray}
The mode-dependent conformal time $\eta_f$ at the end of the
inflation is almost independent of the comoving momentum $k$. Note
that from Eq.~(\ref{k nu:22}),
\begin{eqnarray} \label{mu:nu}
\nu &\simeq & \mu - c h^2, ~~
c\equiv\frac{\mu^2-1/4}{2\mu}\left(w_3-\frac{\alpha_n}{\mu-1}\right)
, \\
3-2\nu &\simeq& 3-2\mu+ 2c h^2=-\frac{2}{n-1}+ 2 ch^2, \nonumber
\end{eqnarray}
where $c$ is a positive number for $n> 1.53$. $3-2\nu$ changes sign
from positive to negative at $\eta_k=\eta_{flat}$, where
\begin{eqnarray} \label{flat}
\eta_{flat}=
\left[\frac{(2-\mu)(2\mu-3)}{cn^2}\right]^{\frac{1}{2\mu-3}}\eta_0 .
\end{eqnarray}
Since $|3-2\nu| \ll 1$ always, the slightly blue-shifted spectrum
changes into the slightly red-shifted spectrum at $\eta=
\eta_{flat}$.

The power spectrum decreases in time since $\mu-\nu$ is positive
even though it is very small.  For example, for $n=14$, we have
$a(\eta_{flat})/a_0 \sim e^{41}$ and $\displaystyle P_k \sim 0.075
\frac{\kappa^2}{M_P^2}$. Since the currently discussed minimal
duration of inflation is roughly $a/a_0 \sim e^{60}$~\cite{giovan},
$\eta_f$ must be later than $\eta_{flat}$ and the size of the
spectrum decreases in time. We will observe the decreasing red
spectrum of the scalar density fluctuation for high comoving
momentum $k>k_m$.

If $n=19$, on the other hand, we have $a(\eta_{flat})/a_0\sim
e^{60}$ and we will observe fully scale invariant power spectrum for
the high comoving momentum modes for minimal inflation.

There are a couple of features in our result (\ref{pS:UV}) which can
be used to resolve the gravitational hierarchy problem, the extreme
weakness of the gravitational wave (the tensor mode fluctuation)
relative to the large scale density fluctuations. The first is the
fact that the power spectrum decreases as $k$ increases for later
time, which is also the case in the commutative space results. The
second is the time-dependence of the power spectrum, which is a new
feature of the present paper, that may describe the difference of
the power spectra between the structure formation time and the
observation time. These features may solve the gravitational
hierarchy problem without imposing any fine tuning on $H$ or on the
non-commutativity scale $\kappa$.

%%%%%%%%%%%%%%%%%%%%%%%%%%%%%%%%%%%%%%%%%%%%%%%%%%
\subsection{Speculation for IR modes}
%%%%%%%%%%%%%%%%%%%%%%%%%%%%%%%%%%%%%%%%%%%%%%%%%%
In the previous subsection, we deferred the discussion on the
evolution of the modes of smaller frequencies than $k_m$. Due to the
definition of $k_m$, (\ref{valid}), these modes always satisfy in
its physical region,
\begin{eqnarray} \label{fir:1}
\frac{k}{a(t) \kappa} \ll 1.
\end{eqnarray}
The Far Infra-Red (FIR) modes are created at $\eta=\eta_k^0<
\eta_0$. For $\eta_k^0<\eta <\eta_m$, the condition~(\ref{fir:1})
can be interpreted as
\begin{eqnarray} \label{k:H}
k^2 \ll a^2 H^2, ~~ \mbox{ for } \eta < \eta_m ,
\end{eqnarray}
since $\eta_m$ is the time for $H/\kappa \sim 1$.

To figure out the dynamics of these modes, we must have reasonable
approximation on the action around $H^2/\kappa^2 \sim 1$. We assume
that the action~(\ref{S:final}) is still valid even though the
explicit functional forms of $m$ and $f$ are not known. We keep
Eq.~(\ref{bmu}) since the equation comes from the requirements of
quantization and unitarity of the time evolution. To show that the
mode solution at time $\eta_m$ is almost independent of $k$, we use
several steps of reasoning. First, we show that $m(k,\eta_k)$ is
very small until $\eta_k <\eta_m$. Next, we argue that $\Omega^2_k$
is also very small. Finally, we argue that $\Omega_k^2$ is almost
independent of $k$ for this region of time. With these, we have the
result that the matching condition of $\varphi_k(\eta_k)$ at $\eta_k
\sim \eta_m$ is almost independent of $k$.

$m(k,\eta)$ is defined by the first order differential
equation~(\ref{bmu}) with $m=1$ at $\kappa t\rightarrow \infty$. The
value of $m$ increases in time asymptotically approaching to, $m\sim
\bar \mu$ at $\eta=0$. Therefore, to get the behavior of $m$, we
need to analyze its behavior around $m\sim 0$. In this case, the
differential equation~(\ref{bmu}) becomes
\begin{eqnarray} \label{dm:0}
\frac{\dot m}{m} \simeq \sqrt{-\frac{3\kappa^2\bar \mu}{\gamma}}.
\end{eqnarray}
Thus, to have a consistent, real-numbered value of $m$, we must have
$ \bar \mu /\gamma<0$ in this region and
\begin{eqnarray} \label{mm}
m(k,\eta_k) \simeq
\left[\Lambda-\int_{\eta_k^0}^{\eta_k}\sqrt{\frac{-3\kappa^2 \bar
\mu a^2}{\gamma}} d\eta_k \right]^{-1} ,
\end{eqnarray}
where the integration constant $\Lambda$ is a large number which
ensures the value of $m$ to be small.  The value of $m$ increases to
$O(1)$ around $\bar \mu /\gamma \sim 0$ since for positive $\bar
\mu$ we have $m \sim \bar \mu$ as a 0$^{\rm th}$ order
approximation. Let $\eta_k^c$ be the value of $\eta_k$ satisfying
$\bar \mu(\eta_k^c)/\gamma(\eta_k^c) =0$. Then, we have $m \sim
1/\Lambda \ll 1$ for $\eta_k \ll \eta_k^c$ and $m \sim \bar \mu$ for
$\eta_k > \eta_k^c$. Rough estimation on the value of $\eta_k^c$ is
possible by using the asymptotic form of $m(k, \eta_k)\sim \bar
\mu$. Since
\begin{eqnarray} \label{..}
\frac{k^2}{\kappa^2 a^2(t_k^c)}\simeq 1- \alpha_n
\frac{H^2(t_k^c)}{\kappa^2} <1,
\end{eqnarray}
where $t_k^c$ is the time corresponding to the mode-dependent
conformal time $\eta_k^c$, and the mode $\phi_k$ is generated before
$\eta_k^c$. Since $H(\eta_m)\sim \kappa$, we must have $\eta_k^c\sim
\eta_m+\cdots $. Therefore, the zeroth order of $\eta_k^c$ is
independent of $k$.

Next, we consider the function $f$ of Eq.~(46) for $\eta_k \ll
\eta_k^c$. From Eqs.~(\ref{bmu}), (\ref{dm:0}), and the change of
variable $\displaystyle \bar f=\frac{\sqrt{-\gamma \bar
\mu}}{\kappa^2 a m} f$, we get
\begin{eqnarray} \label{f:0}
\frac{3\omega}{\kappa^2}+\partial_\eta{\bar f}\simeq -\frac{\bar
f^2}{\gamma \bar \mu}
 \geq 0,
\end{eqnarray}
where we have ignored the term proportional to $m$ since $m$ is very
small. For this differential equation to be well defined, $\bar f$
must be a function of $O(1)$. Thus, $\displaystyle f=\frac{\kappa^2
a m}{\sqrt{-\gamma \bar \mu}} \bar f$ is of the same order as $m$,
and is very small. Therefore, the potential ($\propto m f \propto
1/\Lambda^2$) is very small.

Finally, note that $\Omega_k^2/(a^2\kappa^2)$ is a function of
$k^2/(a^2\kappa^2)$ and $H^2/\kappa^2$. Because of Eq.~(\ref{k:H})
for FIR modes, we guess that $\Omega_k$ may be almost independent of
$k$ for $\eta_k <\eta_k^c$. This implies that the initial mode
solution created at $\eta_k^0$ of the FIR modes are almost
independent of $k$.

Since $m$ is small, small variation $\delta\eta$ corresponds to a
large variation in the mode dependent conformal time $\delta
\eta_k\sim \Lambda\delta\eta$ in this region. Since the
multiplication of potential and the time-interval,
\begin{eqnarray} \label{..dphi}
\frac{\delta \phi_k'}{\phi_k}\sim \Omega_k^2 \delta \eta_k \sim
\frac{1}{\Lambda},
\end{eqnarray}
is small, the mode solution does not change much until $\eta_k^c$
after the mode generation at $\eta_k^0$. Then, the state at
$\eta_k=\eta_k^c$ is not much different from its ground state at
$\eta_k^0$.
\begin{figure}[htbp]
\begin{center}
\includegraphics[width=.6\linewidth,origin=tl]{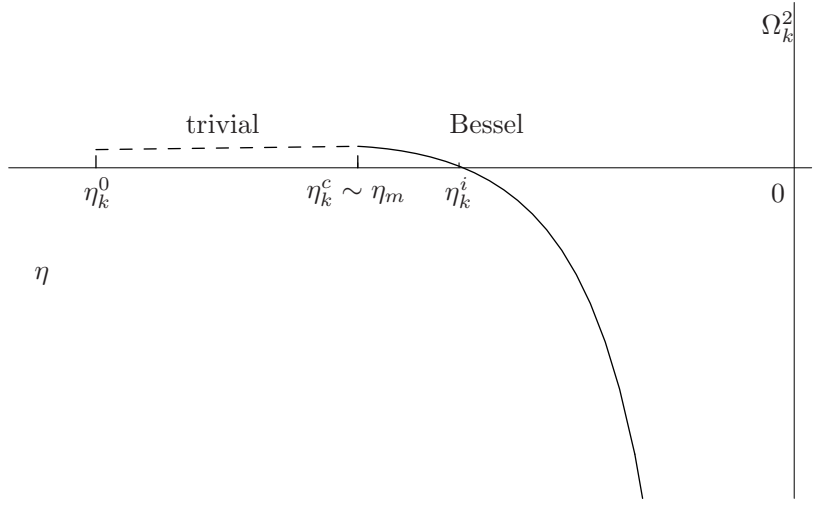}\hfill%
\end{center}
\caption{Schematic plot of $\Omega_k^2$ for IR modes. .}
\label{o2:fig3}
\end{figure}

Collecting all of the above arguments, we propose that the initial
mode solutions, which are almost independent of $k$, vary only by a
small amount during $\eta_k^0<\eta<\eta_k^c\sim \eta_m$. The
evolution for $\eta_k>\eta_k^c$ will be described by the same method
as that of the UV modes with the matching condition at $\eta_k^c$.
The mode solution is given by Eq.~(\ref{v:0}) and its coefficients
are given by Eq.~(\ref{AB:L}). For these modes we have $-k \eta_k^c(
\sim  -k \eta_m) <n(=- k_m \eta_m )$. In the case of a FIR modes, we
have $|k \eta_k^c| \ll 1$. Then, we can use the form~(\ref{sol:0})
of the Bessel function to determine $B_k$:
\begin{eqnarray} \label{B:IR}
B_k \sim -\frac{\pi(\nu+1/2)}{2^{\nu+3/2}\nu
\nu!}\sqrt{\frac{k}{\Omega_k(\eta_k^c)}} (-k \eta_k^c)^{\nu-1/2} ,
\end{eqnarray}
where we have assumed $\eta_k^c\Omega_k(\eta_k^c)\ll 1$.

Thus the power spectrum for the FIR modes is given by
\begin{eqnarray} \label{FIR}
P_{FIR}(k) \simeq \frac{n(\nu+1/2)^2}{32\pi^2\nu^2}
\frac{k^3}{a^2(\eta_m) M_P^2\Omega_k(\eta_m)}
\left(\frac{\eta_k}{\eta_k^c}\right)^{2(\mu-\nu)} .
\end{eqnarray}
Since $\eta_k^c \sim \eta_m$, we have strong (proportional to $k^3$)
blue spectrum for this FIR modes. Since $\mu-\nu$ is a small
positive number, the size of the spectrum slowly decreases in
$\eta_k$. It was argued that, in relation to the pre-big-bang
scenario, this kinds of cutoff of power spectrum in the low
frequency region can explain the low CMB quadrupole
moments~\cite{piao}.

For the spectrum to be continuous, the red spectrum in the UV region
must be continuously deformed to the blue spectrum in the FIR
region. Therefore, the spectrum for $a_0 \kappa < k < k_m$ must be
continuously deformed from the weak red spectrum to flat and then to
weak blue spectrum.

%%%%%%%%%%%%%%%%%%%%%%%%%%%%%%%%%%%%%%%%%%
\section{Summary and discussion}
%%%%%%%%%%%%%%%%%%%%%%%%%%%%%%%%%%%%%%%%%%%%

We have studied the efffects of the $\kappa-$deformation of
Robertson-Walker space on the evolution of metric fluctuations in
expanding cosmological background. For a given noncommutative
$\kappa-$deformed inflationary universe the cosmological background
is still described by the Einstein equation since the background
fields only depend on one variable $t$ so that the homogeneity and
isotropy of the Robertson-Walker space are kept. The equation for
linear fluctuations, however, are modified. We have shown that the
modification takes the form of nonlocal interaction of the
fluctuating field with itself and with the background. We have
analyzed the system by perturbatively expanding the action up to the
first order $H^2/\kappa^2$.

An important consequence of the space-time non-commutativity is that
for each wave number $k$, there exists an earliest time $\eta_k^0$
at which the fluctuating mode is created. The origin of the mode
generation phenomena is a direct consequence of the
$\kappa-$deformation which introduces an upper bound of the comoving
wave-number by Eqs.~(\ref{klimit}) and (\ref{k:max1}). We assume
that the fluctuation starts out with its vacuum amplitude at
$\eta_k^0$ since the number of excitations should be conserved
during the creation process. Moreover, this condition restricts the
physical frequency to be smaller than $\kappa$. This condition
determines the initial condition of given modes for a given initial
time. The deformation also generates the correction terms
proportional to $H^2/\kappa^2$ and $k^2/(\kappa^2a^2)$ to the
frequency squared, which determines the time evolution of mode $k$.

There are two main results for the corrections to the power spectrum
of the metric fluctuation due to the deformation. The first is that
the deformation alters both the time dependence and the momentum
dependence of the power spectrum. Especially, in the case of a power
law inflation, we have shown that the power spectrum slowly
decreases in time as $(\eta_k/\eta_0)^{2(\mu-\nu)}$, where
$\mu-\nu\simeq c h^2$ is a very small but positive. This time
dependence of the power spectrum is a new feature of the present
approach in contrast to the result of the commutative space case and
to the result of Ref.~\cite{ho}. The fact that the power spectrum
decreases in time in addition to the existence of the red shifted
spectrum, can be used to resolve the gravitational hierarchy problem
for the tensor mode fluctuation. Another consequence of the
deformation is that the momentum dependence of the power spectrum
for UV modes is also dependent on time as $k^{3-2\nu}$, where $\nu$
is given by Eq.~(92). Note that $3-2\nu$ is a small positive number
for $\eta_f< \eta_{flat}$ and is a small negative number for
$\eta_f>\eta_{flat}$. Therefore, there is a period ($\eta <
\eta_{flat}$) of blue spectrum in the earlier time of the inflation
and the spectrum becomes red after the time $\eta_{flat}$, which is
determined by $n$. Maximal red shift occurs to the power spectrum as
$\eta_f \rightarrow \infty$ in which case $\mu=\nu$. Another
interesting effect appears in the power spectrum of infra-red modes.
Although we cannot obtain the dynamics of infra-red modes
explicitly, we have suggested a form of the power spectrum from the
consistency requirements. The power spectrum of the far infra-red
modes ($k< a_0 \kappa$), with $a_0$ the scale factor at
$t=1/\kappa$, have a cutoff proportional to $k^3$ even though the
explicit procedure needs much refining since the dynamics at early
times are not known. One can use the existence of this cutoff to
explain the low CMB quadrupole moment. For the spectrum to be
continuous, this $k^3$ type power may change as $k$ increases. We
know that for ultra-violet mode ($k>k_m$, Eq.~(89)) the spectrum
becomes slightly red shifted for $\eta > \eta_{flat}$. Therefore,
the spectrum may change from weak blue to flat spectrum for $ a_0
\kappa < k< k_m$.

Brandenberger and Ho~\cite{ho} computed the effect of the stringy
space-time uncertainty relation to the power spectrum of metric
fluctuations for power law inflation in the Robertson-Walker space.
It is interesting to compare their results with that of the present
paper since we have started from a different commutation
relation~(\ref{com:RW}). We start from the basic commutation
relation~(\ref{com:RW}) and construct the theory from the first
principle, although we have to use the perturbation to compute the
physical effects of the deformation. Let us consider the power
spectra for time $\eta_f>\eta_{flat}$. The spectra of Ref.~\cite{ho}
changes from weak red for ultra-violet modes to weak blue for
infra-red modes. In our case, the far infra-red modes behaves as
$k^3$. This is due to the fact that the far infra-red modes becomes
almost independent of $k$ for the very early times $\eta < \eta_m$.
Since the generation time for the ultra-violet modes is of similar
form as in Ref.~\cite{ho}, the behavior of the spectrum in our case
is similar to theirs for UV modes. The spectrum for $k>a_0 \kappa$
changes from weak blue to weak red as $k$ increases. A totally new
phenomena due to the $\kappa-$deformation, which is absent in the
cases of Ref.~\cite{ho} and the metric fluctuations in the
commutative Robertson-Walker model, is the time-dependence of the
spectra. We have shown that both the spectra of the ultra-violet and
infra-red modes decrease slowly in time.

We have used the perturbative approximation to obtain the generic
feature of the physical effect of the $\kappa-$deformation on the
cosmological evolution. It would be interesting if one could develop
some nonperturbative approximation methods to extract better
information from the nonlocal theory described by the
action~(\ref{S:nl}).

\newpage

\vspace{.5cm}
\begin{acknowledgments}
This work was supported in part by Korea Research Foundation under
Project number KRF-2003-005-C00010 (H.-C.K. and J.H.Y.) and by Korea
Science and Engineering Foundation Grant number
R01-2004-000-10526-0(CR).
\end{acknowledgments}
%\vspace{.2cm}

%\begin{appendix}

%\end{appendix}%

%\bibliographystyle{unsrt}
%\bibliography{bibli3}

\vspace{4cm}

%%%%%%%%%%%%%%%%%%%%%%

\end{document}